\documentclass[12pt,twoside]{article}

\usepackage{amsmath,amsthm,amssymb,euscript,amscd,graphicx, color, amsfonts, fleqn }

\usepackage{graphicx}

\definecolor{mybackgroundcolor}{rgb}{1,0,1}
\definecolor{plum}{rgb}{.5,0,.5}
\definecolor{dkred}{rgb}{.5,0,0}
\definecolor{ddkred}{rgb}{.35,0,0}
\definecolor{dkblue}{rgb}{.1,0,.6}
\definecolor{ddkblue}{rgb}{0,0,.25}
\definecolor{dkgreen}{rgb}{0,.5,0}
\definecolor{ddkgreen}{rgb}{0,.35,0}
\definecolor{dkgr2}{rgb}{0,.57,0}
\definecolor{dkgr3}{rgb}{0,.64,0}
\definecolor{dkgr4}{rgb}{0,.71,0}

\pagecolor{white}

\newcommand{\be}{\begin{equation}}
\newcommand{\ee}{\end{equation}}
\newcommand{\bd}{\begin{displaymath}}
\newcommand{\ed}{\end{displaymath}}
\newcommand{\bea}{\begin{eqnarray}}
\newcommand{\eea}{\end{eqnarray}}
\newcommand{\rhp}{\mathbb{C}_{+}}
\newcommand{\Hi} {{\cal H}^\infty}

\begin{document}
\color{ddkred}

\begin{center}\textbf{ {\Large  Control of Systems with Infinitely Many Unstable Modes and Strongly Stabilizing Controllers Achieving a Desired Sensitivity}\\[2ex]
 Suat G\"um\"u\c{s}soy and Hitay \"Ozbay \\
 Collaborative Center of Control Science\\ Department of Electrical
Engineering
\\ The Ohio State University \\ 2015 Neil Avenue, Columbus, OH
43210\\[2ex] \color{ddkgreen} {\it MTNS 2002}}

\end{center}

\color{dkblue}
 
\begin{abstract}

\color{black}
In this paper we consider a class of
linear time invariant systems with infinitely many unstable modes.
By using the parameterization of all stabilizing controllers, we
show that $\Hi$ controllers for such systems can be computed using
the techniques developed earlier for infinite dimensional plants
with finitely many unstable modes. We illustrate connections
between the problem solved here and an indirect method for
strongly stabilizing $\Hi$ controller design for systems with time
delays.
\end{abstract}

\section{Introduction}
\setcounter{equation}{0}

\color{black} If a stable controller results in a stable feedback
system, then it is said to be a strongly stabilizing controller,
\cite{V85}. There are many practical applications where strongly
stabilizing $\Hi$ controllers are desired, see e.g.
\cite{Sideris,Barabanov,Jacobus,Ito,Ganesh,ZO99,ZO00,Zhou01} and
their references.  These papers on $\Hi$ strong stabilization deal
with direct design methods for finite dimensional plants. The
problem is still an open research area for infinite dimensional
plants.

\begin{figure}[ht]
\begin{center}
\includegraphics[width=0.6\linewidth]{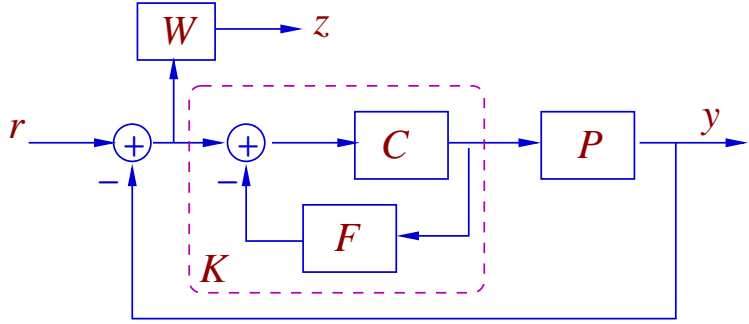}

\color{ddkred} \caption{\it (i) $F=0$: the weighted  sensitivity is
$\Hi$ optimal; (ii) $F\ne 0$: the controller $K$ is stable.}
\label{SystemFig}
\end{center}
\end{figure}

It is known that for a certain class of time delay systems the
optimal $\Hi$ controllers designed for sensitivity minimization
lead to controllers with infinitely many unstable modes,
\cite{Flamm,Lenz}. An indirect way to obtain a strongly
stabilizing controller, in this case, is to internally stabilize
the optimal sensitivity minimizing $\Hi$ controller, while keeping
the sensitivity deviation from the optimum within a desired bound.
The proposed scheme is illustrated in Figure~\ref{SystemFig}: the
objective is to have a stable feedback system, and to minimize the
weighted sensitivity function $W S:=W (1+PK)^{-1}$, with a stable
$K$. We will assume that for given $W$ and $P$, the optimal $\Hi$
controller $C_{\rm opt}$ is determined. Then, $F$ will be designed
to yield a stable $K$, such that the feedback system remains
stable, and $W S$ is ``relatively close'' to the optimal weighted
sensitivity $W S_{\rm opt}:=W (1+PC_{\rm opt})^{-1}$. See
Section~3 for more precise definition of this problem.

\color{black}

 When the plant, $P$, contains a time delay, and the
sensitivity weight $W$ is bi-proper, the indirect approach
outlined above requires internal stabilization of $C$ (which
contains infinitely many unstable modes) by $F$. In the next
section we will see a solution to the two-block $\Hi$ control
problem involving a plant with infinitely many poles in the open
right half plane. Then, the results of Section~2 will be used in
Section~3 to derive sufficient conditions for solvability of the
stable $\Hi$ controller design problem considered here for systems
with time delays.
%A numerical example is given in Section~4, and
Concluding remarks are made in Section~4.

\color{dkblue}
\section{$\Hi$ Control of Systems with Infinitely Many Unstable Modes}

\color{black}
 In order to be consistent with the notation used in
the rest of the paper, in this section $P_C$ and $C_F$ denote the
``plant'' and the ``controller'' respectively. In the next section
$P_C$ will be the optimal $\Hi$ sensitivity minimizing controller, and
$C_F$ will be $F$. Assume that
\bd
P_C(s)=\frac{N(s)}{M(s)}
\ed
where $M$ is inner and infinite dimensional (it has infinitely many
zeros in $\rhp$, that are unstable poles of $P_C$), $N=N_{i} \;
N_{o}$ with $N_{i}$ being inner finite dimensional, and
$N_o,N_{o}^{-1}\in \Hi$.

Following the controller parameterization of Smith, \cite{S89}, we
form the Bezout equation, in terms of $X,Y\in \Hi$
\be
N X+M Y=1 \label{Bezout}
\ee
i.e.
\bd
X(s)=\left( \frac{1-M(s)Y(s)}{N_{i}(s)} \right) \; N_{o}^{-1}(s).
\ed
Let $z_{1},...,z_{n}$ be the zeros of $N_{i}(s)$ in $\rhp$, and assume
that they are distinct. Then, there are finitely many interpolation
conditions on $Y(s)$ for $X(s)$ to be stable, i.e.
\bd
Y(z_{i})=\frac{1}{M(z_{i})}.
\ed
Thus by Lagrange interpolation, we can find a finite dimensional $Y \in
\Hi$ and infinite dimensional $X \in \Hi$ satisfying (\ref{Bezout}),
and all controllers stabilizing the feedback system formed by the
plant $P_C$ and the controller $C_F$ are parameterized as follows,
\cite{S89},
\be
C_F(s)=\frac{X(s)+M(s)Q(s)}{Y(s)-N(s)Q(s)} \quad \text{where} \; Q(s)
\in \Hi ~ \text{and} \; (Y(s)-N(s)Q(s))\neq 0
\ee
Note that if our concern is simply stabilization of $P_C$, then we can
select $Q(s)=0$ and $C_F(s)=\frac{X(s)}{Y(s)}$ is a stabilizing
controller. But in the next section we will need to solve the following
two block $\Hi$ problem. First note that,
\bd
(1+P_C(s)C_F(s))^{-1}=M(Y(s)-N(s)Q(s))
\ed
\be
P_C(s) C_F(s)(1+P_C(s)C_F(s))^{-1}=N(s)(X(s)+M(s)Q(s)).
\ee
Then, in terms of the free controller parameter $Q$, we define the
$\Hi$ problem as finding
\be \label{twobl}
\inf_{C_F~{\rm stabilizies}~P_C}
\left\| \left[\begin{array}{c}
W_{1}(1+P_CC_F)^{-1} \\
W_{2} P_C C_F(1+P_CC_F)^{-1} \end{array} \right] \right\|_{\infty} =
\inf_{Q\in \Hi} \left\| \left[\begin{array}{c}
W_{1}(Y-N Q) \\
W_{2}N(X+M Q) \end{array} \right] \right\|_{\infty}.
\ee
where $W_1$ and $W_2$ are given finite dimensional (rational) weights. By using the Bezout equation, we can define
\begin{eqnarray}
\gamma(Q) &:=& \left\| \left[\begin{array}{c}
W_{1}Y-W_{1}NQ \\
W_{2} N\left( \frac{1-MY}{N} \right) +W_{2}MNQ  \end{array} \right]
 \right\|_{\infty} \\
& = & \left\| \left[\begin{array}{c}
W_{1}Y-W_{1}N_{i}(N_{o}Q) \\ \nonumber
W_{2} (1-MY) +W_{2}M N Q)
\end{array} \right] \right\|_{\infty} \\
& = & \left\| \left[\begin{array}{c}
W_{1}(Y-N_{i}(N_{o}Q)) \\ \nonumber
W_{2}(1-M(Y-N_{i}(N_{o}Q)))  \end{array} \right] \right\|_{\infty}
\end{eqnarray}
In summary, the $\Hi$ optimization problem reduces to
\begin{equation} \label{eq:prob1}
\inf_{Q \in \Hi}\gamma(Q)=\inf_{Q_{1} \in \Hi} \left\|
\left[\begin{array}{c} W_{1}(Y-N_{i}Q_{1}) \\
W_{2}(1-M(Y-N_{i}Q_{1}))  \end{array} \right]
\right\|_{\infty}
\end{equation}
where $Q_{1}=N_{o}Q$, and note that $W_{1}(s), W_{2}(s),
N_{i}(s), Y(s)$ are rational functions, and $M(s)$ is inner infinite
dimensional.

The problem defined in (\ref{eq:prob1}) has  the same structure as the
problem dealt in Chapter~5 of the book \cite{FOT} (by Foias, \"Ozbay
and Tannenbaum, (F\"OT)), where skew Toeplitz approach has been used for
computing $\Hi$ optimal controllers for infinite dimensional systems
with finitely many poles in $\rhp$. Our case is the dual of the
problem solved in \cite{FOT}, that is there are infinitely many poles
in $\rhp$, but the number of zeros in $\rhp$ is finite. Thus by
mapping the variables as shown below, we can use the results of
\cite{FOT} to solve our problem:
\bd
W_{1}^{F\ddot{O}T}(s)=W_{2}(s)
\ed
\bd
W_{2}^{F\ddot{O}T}(s)=W_{1}(s)
\ed
\bd
X^{F\ddot{O}T}(s)=Y(s)
\ed
\bd
M_{d}^{F\ddot{O}T}=N_{i}(s)
\ed
\bd
M_{n}^{F\ddot{O}T}(s)=M(s)
\ed
\bd
N_{o}^{F\ddot{O}T}(s)=N_{o}(s).
\ed

If we consider the one block problem only, with $W_2=0$, then
the minimization of $\| W_{1}(Y-N_{i}Q_{1}) \|_{\infty}$  is simply a
finite dimensional problem. On the other hand, the one block problem
obtained by putting $W_1=0$, i.e. minimizing $\|
W_{2}(1-M(Y-N_{i}Q_{1})) \|_\infty$ over $Q_1 \in \Hi$, is
an infinite dimensional problem.

%\newpage
\color{dkblue}

\section{Stable $\Hi$ Controllers for Delay Systems:
Suboptimal Sensitivity}

\color{black}

 In this section we investigate the indirect method
of obtaining a strongly stabilizing controller for systems with
time delays, subject to a bound on the deviation of the
sensitivity from its optimal value. First we examine the optimal
sensitivity problem for stable delay systems and illustrate that
the corresponding optimal controller has the structure of $P_C$
introduced in Section~2.

\color{dkblue}
\subsection{Optimal Sensitivity Problem for Delay Systems}

\color{black}
 Consider the feedback system shown in
Figure~\ref{SystemFig}, where $P(s)=e^{-hs}N_{p}(s)$ and
$W(s)=\frac{1+\alpha s}{s+\beta}$. We assume that
$N_{p}$,$N_{p}^{-1}\in\Hi$. By using the method developed in
\cite{FOT,TO95}, we calculate the optimal controller,
$C_{opt}(s)$, minimizing the weighted sensitivity $W(1+PC)^{-1}$
over all stabilizing controllers, as follows.
%First define
%\begin{eqnarray}
%E_{\gamma}(s)&=&\left( \frac{W(s)W(-s)}{\gamma^2}-1
%\right)=\frac{(1-\gamma^2\beta^2)+(\gamma^2-\alpha^2)s^2}{\gamma^2(\beta^2-s^2)}\\
%F_{\gamma}(s)&=&\frac{\gamma}{W(s)}\left(\frac{\beta-s}{\beta+s}\right)=\gamma
%\left( \frac{\beta-s}{1+\alpha s} \right)
%\end{eqnarray}
The smallest $\gamma$  satisfying the phase equation given below is
the optimal (smallest achievable) sensitivity level:
\begin{equation} \label{eq:pheqn}
h\omega_{\gamma}+\tan^{-1}\alpha\omega_{\gamma}+
\tan^{-1}\frac{\omega_{\gamma}}{\beta}=\pi
\end{equation}
where $\omega_{\gamma}=\sqrt{
  \frac{1-\gamma^{2}\beta^{2}}{\gamma^{2}-\alpha^{2}}}$,
  and $\alpha<\gamma<\frac{1}{\beta}$.
Once $\gamma_{opt}$ is computed as above, the corresponding optimal
controller is
\begin{equation} \label{eq:Copt}
C_{opt}(s)=\frac{(1-\gamma_{opt}^2\beta^2)+
(\gamma_{opt}^2-\alpha^2)s^2}{\gamma_{opt}(\beta+s)(1+\alpha s)}
\quad \frac{N_{p}^{-1}(s)}{1+\gamma_{opt} \left(
\frac{\beta-s}{1+\alpha s} \right) e^{-hs}}.
\end{equation}

%to calculate optimal controller,

%\be
%C_{opt}(s)=E_{\gamma}(s)m_d(s)
%\frac{N_{p}^{-1}F_{\gamma}(s)L(s)}{1+m_n(s)F_{\gamma}(s)L(s)}
%\ee

%\noindent
%L(s) take values $\pm1$,and after substituting $E_{\gamma}(s)$ and $F_{\gamma}(s)$

%\be
%C_{opt}(s)=\frac{(1-\gamma^2\beta^2)+(\gamma^2-\alpha^2)s^2}
%{\gamma(\beta+s)(1+\alpha s)} \quad
%\frac{N_{0}^{-1}\pm1}{1+\gamma \left( \frac{\beta-s}{1+\alpha s} \right) \pm1 e^{-hs}}
%\ee

%\noindent
%By checking magnitude and phase conditions for $\gamma_{opt}$, we
%can see that maximum $\gamma$ value takes place when $L(s)=1$.
%Therefore $C_{opt}(s)$ is

\noindent Also, define the optimal sensitivity function  as
$S_{opt}(s)=(1+P(s)C_{opt}(s))^{-1}$, then,
\begin{equation} \label{eq:Sjw}
S_{opt}(j\omega)=
\frac{1+\left( \frac{\gamma_{opt} (\beta-j\omega)}{1+\alpha j\omega} \right)
e^{-jh\omega}}
{1+\left( \frac{1-\alpha j\omega}{\gamma_{opt} (\beta+j\omega)} \right)
 e^{-jh\omega}}.
\end{equation}
In \cite{Flamm}, it was mentioned that $\Hi$-optimal controllers may
have infinitely many right half plane poles. Here we will give a proof
based on elementary Nyquist theory: if $S_{opt}^{-1}(j\omega)$
encircles the origin infinitely many times, we can say that
$C_{opt}(s)$ has infinitely many right hand poles, because $P(s)$ does
not have any right half plane poles. For $s=j\omega$ as
$\omega\rightarrow\infty$, we have
\bd
S_{opt}^{-1}(j\omega) \rightarrow
\frac{1-\frac{\alpha}{\gamma_{opt}}e^{-jh\omega}}
{1-\frac{\gamma_{opt}}{\alpha}e^{-jh\omega}}
\ed
and $|S_{opt}^{-1}(j\omega)| \rightarrow \frac{\alpha}{\gamma_{opt}}$.
Since $\alpha<\gamma_{opt}<\frac{1}{\beta}$, we can say that
$|S_{opt}^{-1}(j\omega)|$ has constant magnitude between 0 and 1 for
sufficiently large $\omega$. For $\omega_k=\frac{2\pi k}{h}$, as
$k\rightarrow \infty$ the phase of $S_{opt}^{-1}(j\omega_k)$ tends to
$-\pi$. In other words, $S_{opt}^{-1}(j\omega)$ intersects negative
part of the real axis near $\omega_k=\frac{2\pi k}{h}$, as
$k\rightarrow \infty$. Similarly, $S_{opt}^{-1}(j\omega)$ intersects
positive part of the real axis near $\omega_k=\frac{(2k+1)\pi}{h}$ as
$k\rightarrow \infty$. Thus $S_{opt}^{-1}(j\omega)$ encircles the
origin infinitely many times, which means that $C_{opt}(s)$ has
infinitely many poles in $\rhp$.

\noindent {\bf Remark}. Let
$m_1(j\omega)=\left( \frac{\beta-j\omega}{\beta+j\omega} \right) e^{-jh\omega}$,
$m_2(j\omega)=\left( \frac{1-\alpha j\omega}{1+\alpha j\omega} \right) e^{-jh\omega}$
and $g(j\omega)=\gamma_{opt} \left( \frac{\beta+j\omega}{1+\alpha j\omega} \right)$.
%Note that $|m_i(j\omega)|=1$ for $i=1,2$. \\
Then,
\bd
W(j\omega)S_{opt}(j\omega)= \gamma_{opt} \left( \frac{g^{-1}(j\omega) +
m_1(j\omega)}{1+ g^{-1}(j\omega)m_2(j\omega)} \right)=\gamma_{opt} \left(
\frac{1 +g(j\omega)m_1(j\omega)}{g(j\omega)+m_2(j\omega)} \right)
\ed
and hence $|W(j\omega)S_{opt}(j\omega)|=\gamma_{opt}$ as expected.

\color{dkblue}
\subsection{Sensitivity Deviation Problem}
\color{black}

Recall that the $\Hi$ optimal performance level was defined as
\bd
\gamma_{0}:=\gamma_{opt}=\inf_{ C~{\rm stab.}~ P }
 \| W(1+PC)^{-1}\|_{\infty}
\ed
where $W(s)=\frac{1+\alpha s}{s+\beta}$, with $\alpha>0$,
$\beta>0$, $\alpha \beta<1$, and $P(s)=N_{p}(s)M_{p}(s)$, with
$N_{p},N_{p}^{-1}\in \Hi$, and $M_{p}$ is inner and infinite
dimensional, e.g. $M_p(s)=e^{-hs}$. We have obtained the optimal
controller for the sensitivity minimization problem in
(\ref{eq:Copt}).

\noindent {\bf Claim}: The optimal $\Hi$ controller is in the form
\be
C_{opt}(s)=\frac{N_{p}^{-1}(s)N_{c}(s)}{D_{c}(s)}
\label{CoptStructure}
\ee
where $D_{c}$ is inner infinite dimensional and $N_{c},
N_{c}^{-1}\in\Hi$.

It is easy to verify this claim by comparing (\ref{eq:Copt}) with
(\ref{CoptStructure}): we see that
\be \label{eq:Ncs}
N_c(s)=\frac{\gamma_{opt}^{-2}W^{2}(s)m_2(s)-m_1(s)}
{1+\gamma_{opt}^{-1}W(s)m_2(s)}=
\frac{1}{\gamma_{opt}^2(\beta+s)^2} \;
\frac{(1-\gamma_{opt}^2\beta^2)+(\gamma_{opt}^2-\alpha^2)s^2}
{1+\gamma_{opt}^{-1}\left(
    \frac{1-\alpha s}{\beta+s} \right)e^{-hs}}
\ee
\begin{eqnarray} \label{eq:Dcs}
D_c(s)&=&\frac{\gamma_{opt}^{-1}W(s)+m_1(s)}
{1+\gamma_{opt}^{-1}W(s)m_2(s)}= \gamma_{opt}^{-1}W(s) \left(
\frac{1+m_1(s)\gamma_{opt}W^{-1}(s)} {1+m_2(s)\gamma_{opt}^{-1}W(s)}
\right) \nonumber \\
&=&\gamma_{opt}^{-1}W(s)(1+P(s)C_{opt}(s))^{-1}= D_{c}(s)=
\frac{\left( \frac{\beta-s}{\beta+s}
\right)e^{-hs}+\gamma_{0}^{-1} \left( \frac{1+\alpha
    s}{\beta +s} \right)}{1+\gamma_{0}^{-1} \left( \frac{1-\alpha
    s}{\beta+s} \right)e^{-hs}}
\end{eqnarray}
where $m_1(s)=\left( \frac{\beta-s}{\beta+s} \right) e^{-hs}$,
$m_2(s)=\left( \frac{1-\alpha    s}{1+\alpha s} \right) e^{-hs}$. Note
that $N_c(s)$ has no right half poles or zeros (it has only two
imaginary axis poles that are cancelled by the zeros at the same
locations). Therefore $N_c,N_c^{-1}\in\Hi$. Also, it is easy to check
that $D_c$ is inner and infinite dimensional.

Note that,
\bd
D_c=\gamma_{0}^{-1}WS_{0}=
\gamma_{0}^{-1}W(1+PC_{opt})^{-1}
=\left(
  \gamma_{0}^{-1}W
  \frac{D_{c}}{D_{c}+M_{p}N_{c}}
   \right) .
\ed

Our goal is to have a stable controller $K$, by an appropriate
selection of $F$:
\bd
K(s)=\frac{C_{opt}(s)}{1+F(s)C_{opt}(s)} .
\ed
At the same time we would like to have the resulting sensitivity
function,
\be
S(s)=(1+P(s)K(s))^{-1}=
\left(1+M_{p}(s)N_{p}(s)\frac{N_{p}^{-1}(s)\frac{N_{c}(s)}
{D_{c}(s)}}{1+F(s)N_{p}^{-1}(s)\frac{N_{c}(s)}{D_{c}(s)}}
\right)^{-1} ,
\ee
to be close to the optimal sensitivity, $S_{opt}=(1+PC_{opt})^{-1}$. By the
parameterization of the set of all stabilizing controllers for
$C_{opt}$ \cite{S89}, $F$ can be written as,
\bd
F(s)=\frac{X(s)+D_{c}(s)Q(s)}{Y(s)-N_{p}^{-1}(s)N_{c}(s)Q(s)}
\ed
with $N_{p}^{-1}(s)N_{c}(s)X(s)+D_{c}(s)Y(s)=1$ which can be solved as
$Y=0$ and $X=N_{c}^{-1}N_{p}$ where $Q\in\Hi$, $Q(s)\neq0$. Then, in
terms of the design parameter $Q$, the functions  $F(s)$, $K(s)$ and
$S(s)$ can be re-written as,
\be \label{eq:Fs}
F(s)=-\frac{N_{c}^{-1}(s)N_{p}(s)+D_{c}(s)Q(s)}{N_{p}^{-1}(s)N_{c}(s)Q(s)}
    =-(Q^{-1}(s)+C_{opt}^{-1}(s))
\ee
\be \label{eq:Ks}
K(s)=\frac{C_{opt}(s)}{1-C_{opt}(s)(Q^{-1}(s)+C_{opt}^{-1}(s))}
    =-Q(s)
\ee
\be
S(s)=(1+M_{p}(s)N_{p}(s)(-Q(s)))^{-1}.
\ee
 Also, sensitivity function $S(s)$ should be stable. We
can define the relative deviation of the sensitivity as
$\|W\left(\frac{S_{0}-S}{S}\right)\|_{\infty}$, then minimizing this
deviation over  $Q\in\Hi$, $Q(s)\neq 0$ is equivalent to
\be
\gamma_{1,opt}=\inf_{Q\in\Hi} \left\| W\left(\frac{S_{0}-S}{S} \right)
\right\|_{\infty}=\inf_{Q\in\Hi} \left\| W
  \left(-\frac{(M_{p}N_{c})(1+D_{c}N_pN_c^{-1}Q)}{D_{c}+M_{p}N_{c}} \right)
  \right\|_{\infty} .
\ee

\noindent Note that, $|D_{c}(j\omega)+M_{p}(j\omega)N_{c}(j\omega)|
=|\gamma_0^{-1}W(j\omega)|$ as shown before. Then,

\begin{equation} \label{eq:gam1opt}
\gamma_{1,opt}=\inf_{\widehat{Q}\in\Hi}\|\gamma_{0}N_{c}(1+D_{c}\widehat{Q})\|_{\infty}
\end{equation}
where $\widehat{Q}=N_pN_c^{-1}Q$. For stability of
the feedback system formed by the resulting controller $K$ and the
original plant $P$, we also want the sensitivity function, $S=(1-M_{p}N_{p}Q)^{-1}$ to be stable.
Once the optimal $Q$ is determined from (\ref{eq:gam1opt}), a
sufficient condition for stability of $S$ (and hence the original
feedback system) can be determined as
\be
|N_p(j\omega )| ~<~ |Q(j\omega)|^{-1} ~~\forall~\omega
\label{SuffCondStab}
\ee

Note that problem defined in (\ref{eq:gam1opt}) is equivalent to a
sensitivity minimization with an infinite dimensional ``weight''
$\gamma_0N_c$ for a stable infinite dimensional ``plant'' $D_c$. For the case where both the
plant and the weight are infinite dimensional, sensitivity
minimization problem is difficult to solve. So, we propose to
approximate the weight by a finite dimensional upper bound function:
find a stable rational weight $W_{1}$  such that
$|\gamma_{0}N_{c}(j\omega)|\leq |W_{1}(j\omega)|$. We suggest
an envelope which is in the form,
\bd
W_{1}(s) =\gamma_{0}~K~\frac{s+\alpha_{1}}{s+\beta_{1}},
\ed
where
\begin{eqnarray*}
K & = & 1+\alpha \gamma_{opt}^{-1}\\
\beta_1 & = & \beta (\gamma_{opt}+\alpha)
 (1-\gamma_{opt}\beta)^{-1} \alpha_1
\end{eqnarray*}
and $\alpha_1$ is determined in some optimal fashion, the details are
in the full version of the paper, \cite{GO2002}.

Then, we can solve the one following block problem as in Section~3.1
\be
\gamma_{1,opt}\leq\gamma_{2,opt}=
\inf_{\widehat{Q}\in\Hi}\|W_{1}(1+D_{c}\widehat{Q})\|_{\infty} .
\ee
Note that $\gamma_{2,opt}$ is the smallest value of $\gamma_2$, in
the range $\gamma_0 K< \gamma_2<(\gamma_0
    K)\frac{\alpha_1}{\beta_1}$, satisfying
\bea \label{eq:pheqn2}
\pi &=&\tan^{-1}\left( \frac{\omega}{\alpha_1}\right)+\tan^{-1}\left(
  \frac{\omega}{\beta_1}\right)+h\omega+\tan^{-1} \left(
  \frac{\gamma_0^{-1} \alpha \omega \cos(h\omega)+(\omega-\gamma_0^{-1}
    \sin(h\omega))}{(\beta+\gamma_0^{-1} \cos(h\omega))+\alpha
    \gamma_0^{-1} \omega \sin(h\omega)} \right) \nonumber \\
& &
-\tan^{-1} \left(\frac{\gamma_0^{-1} \alpha \omega \cos(h\omega)
-(\omega-\gamma_0^{-1}
    \sin(h\omega))}{(\beta+\gamma_0^{-1} \cos(h\omega))-\alpha
    \gamma_0^{-1} \omega \sin(h\omega)} \right)
\eea
where $\omega=\sqrt{\frac{(\gamma_o
    K)^2\alpha_1^2-\gamma_2^2\beta_1^2}
    {\gamma_2^2-(\gamma_0 K)^2}}$.
After finding $\gamma_{2,opt}$, we
    can write the $C_{2,opt}$ as,
\be \label{eq:C2opt}
C_{2,opt}(s)=A(s)\frac{1}{1-D_c(s)B(s)}
\ee
where,
\bea
 \nonumber
 A(s)&=&\frac{(\gamma_{0}^{2}K^2\alpha_{1}^{2}-
 \gamma_{2,opt}^{2}\beta_{1}^{2})+
 (\gamma_{2,opt}^{2}-\gamma_{0}^{2}K^2)s^{2}}
 {\gamma_{0}K\gamma_{2,opt}(\beta_{1}+s)(\alpha_{1}+s)}
 \\
\nonumber B(s)&=&\left( \frac{\gamma_{2,opt}}{\gamma_0 K} \right)
 \left(\frac{\beta_1-s}{\alpha_1+s} \right) .
\eea
In order to calculate $\widehat{Q}_{2,opt}(s)$ corresponding to $C_{2,opt}(s)$,
we will use the transformation
\bd
\widehat{Q}_{2,opt}(s)=\frac{C_{2,opt}(s)}{1+P(s)C_{2,opt}(s)}.
\ed
That gives
\bea \label{eq:Q2opt}
\nonumber \widehat{Q}_{2,opt}(s)&=&A(s)\frac{1}{1-D_c(s)B^{-1}(-s)} \\ &=&
\frac{(\gamma_{0}^{2}K^2\alpha_{1}^{2}-
\gamma_{2,opt}^{2}\beta_{1}^{2})+
(\gamma_{2,opt}^{2}-\gamma_{0}^{2}K^2)s^{2}}
{\gamma_{0}K\gamma_{2,opt}(\beta_{1}+s)(\alpha_{1}s)}~~
\frac{1}{1-D_{c}(s)\left(\frac{\gamma_{0}K}{\gamma_{2,opt}}\right)
\left( \frac{\alpha_{1}-s}{\beta_{1}+s} \right)}
\eea
After finding $\widehat{Q}_{2,opt}(s)$, $F(s)$ can be calculated via
(\ref{eq:Fs}),
\bd
F(s)=-(\widehat{Q}_{2,opt}^{-1}(s)+C_{opt}^{-1}(s))
\ed
where $\widehat{Q}_{2,opt}(s)$ and $C_{opt}(s)$ are found in
(\ref{eq:Q2opt}) and (\ref{eq:Copt}) respectively.\\

Similarly, the resulting  controller $K(s)$ is determined as
\bd
K(s)=-\widehat{Q}_{2,opt}(s)
\ed
which is shown in (\ref{eq:Ks}).

Recall that the largest value of $|N_p(j\omega)|$, for which $K$
becomes a strongly stabilizing controller for $P=M_pN_p$, is
\bd
|N_p(j\omega)| < |K(j\omega )|^{-1} .
\ed
It is also possible to blend this condition with the largest allowable
sensitivity deviation condition. That would result in a two block
$\Hi$ problem (which is slightly more difficult to solve by hand
calculations that are similar to those we have done in this section). We refer
to the full version of the paper, \cite{GO2002}, for the details and a
numerical example.

\newpage
\color{dkblue}

\section{Conclusions}
\color{black}

 In this paper we have considered $\Hi$ control of a class
of systems with infinitely many right half plane poles. We have
demonstrated that the problem can be solved by using the existing
$\Hi$ control techniques for infinite dimensional systems with
finitely many right half plane poles. Connections  with strong
stabilization are made, and we have seen an indirect design method for
stable controllers achieving a desired sensitivity, for infinite
dimensional plants (in particular systems with time delays). There are
alternative direct methods of designing $F$, or an appropriate $K$.
Comparisons of different design methods will be made with examples in
the full version of our paper.

\color{dkblue}

\end{document}